\documentclass{PoS}

\title{Volume independence of large-N QCD with adjoint fermions}

\ShortTitle{Volume independence of large-N QCD with adjoint fermions}

\author{Barak Bringoltz and \speaker{Stephen R. Sharpe}\\ 
        Physics Department University of Washington Seattle, WA 98195-1560\\
        E-mail: \email{barak@phys.washington.edu},
                \email{sharpe@phys.washington.edu}}

\abstract{
It has been proposed that four-dimensional QCD 
with fermions in the adjoint representation
exhibits volume-independence in the large-$N$ limit. 
If correct, this would mean that results for physical
quantities could be obtained from the single-site version of the model.
A necessary condition for volume-independence
is that the $(Z_N)^4$ center-symmetry of the single-site theory is unbroken.
We explore the phase diagram of the theory with a single Dirac fermion
using Wilson fermions for a number of
colors in the range $8 \le N \le 15$, and identify the region in the
parameter space of quark mass and gauge coupling where the symmetry
appears to be unbroken.
Our evidence suggests that this region includes both light and  
heavy quarks, and our results are consistent with this region
extending to the continuum limit. 
}

\FullConference{The XXVII International Symposium on Lattice Field Theory\\
                 July 26-31, 2009\\
                 Peking University, Beijing, China}

\begin{document}

\section{Introduction}
Long ago, Eguchi and Kawai (EK) proposed the striking idea that the properties of
gauge theories
become independent of the volume when the number of colors, $N$, is sent
to infinity~\cite{EK}. 
Using a lattice regularization, this would allow one to determine
physical, infinite-volume, quantities from calculations on a single lattice site,
with both color and space-time degrees of freedom packed into the four link
matrices. This idea is of both theoretical interest---can such matrix models
indeed represent field theories?---and of possible practical utility---it might
allow the determination of large-$N$ properties in a more computationally
efficient way. This would in turn allow one to connect to approximate analytic
approaches to large-$N$ QCD, as well as to learn about QCD itself.

The application of this ``volume-reduction'' to pure gauge theories turned out
not to be simple. As discussed below, the $Z_N^4$ center symmetry of the
$SU(N)$ volume-reduced lattice theory must be unbroken for the equivalence to
infinite volume to hold, but this symmetry is, in fact, spontaneously broken
by the single-site theory~\cite{BHN0,MK0}.
Various proposed ``fixes''---in particular, ``momentum-quenching'' and twisting---have turned out
not to work. See \cite{QEK} for references to this history.
What does work, however,
is to reduce to a lattice whose physical extent exceeds a scale
of about $1\,$fm~\cite{KNN}, and this approach, which, for a lattice spacing $a\approx 0.15$fm, allows one in practice to
work on lattices of size $\sim 6^4$, has been successful in determining
several large-volume physical quantities.

In our present work (summarized here but described in much more detail in
Ref.~\cite{AEK}) we return to the possibility of complete volume reduction
to a single-site theory, but in the context of QCD-like theories with
fermions in two-index representations. In particular, our evidence
suggests that such reduction holds for the theory with ``quarks'' in the
adjoint representation. In addition, it appears that these quarks can be
heavy, with $m \gg \Lambda_{\rm QCD}$, in which case the long distance
theory is the pure-gauge theory. 
In that case the reduction we find 
implements the original idea of EK.

Our motivation comes from the following relations between theories.
Start with QCD, with $N=3$, but consider the fundamental representation
Dirac fermions ($2 N_f$ of them) to be in the equivalent 
antisymmetric (AS) representation with two antiquark indices.
Now take $N\to\infty$ and obtain the ``orientifold'' large-$N$ limit 
of QCD~\cite{CR}---one in which quark-loops are not suppressed.
This theory is ``within $1/N$'' of physical QCD.
Next use the ``orientifold'' equivalence of Armoni, Shifman
and Veneziano~\cite{ASV}, 
which states that the even C 
sector of the large-$N$ theory with $2 N_f$ AS Dirac fermions
is equivalent to the corresponding C-even
sector of the large-$N$ theory with $N_f$ Dirac
fermions in the adjoint irrep.
Finally, follow Kovtun, \"Unsal and Yaffe and
use an orbifold equivalence to relate, when $N\to\infty$, 
the latter field theory to the single-site gauge theory with
$N_f$ adjoint Dirac fermions~\cite{KUY2}.
If these equivalences hold, the large-$N$ single-site theory 
gives results for two interesting
field theories, one of which is close to physical QCD.\footnote{%
For details on how one extracts long-distance quantities
such as the string tension and meson masses from a single-site model
see Refs.~\cite{EK,KNN,AEK,KUY2}.}

In our work to date we have taken $N_f=1$---so that the corresponding
QCD-like theory has two degenerate quarks---but $N_f=1/2$ and $2$ are
also of considerable interest (the former is related to SUSY Yang-Mills
and to 1-flavor QCD, the latter to the potentially nearly-conformal
theory with 2 adjoint Dirac fermions).

A crucial question is, of course, what conditions must be
satisfied for the equivalences to hold. The
necessary and sufficient conditions are as follows~\cite{KUY1}:
\begin{enumerate}
\item
Large-N factorization holds.
\item
The orientifold equivalence
requires that C is not spontaneously broken in either
the theory with AS or that with Adjoint fermions, at large volumes.
\item
The orbifold equivalence requires that translation invariance
is unbroken in the large-volume theory with Adjoint fermions.
\item
The orbifold equivalence also requires that the $Z_N^4$ center
symmetry of the single-site theory with Adjoint fermions be unbroken.
\end{enumerate}
The first three conditions are expected to hold
(although C can be broken for small volumes, and translation 
invariance can break at nonzero density)
and it is the last condition that is the weakest link. Indeed, as noted
above, this is the condition that fails in the pure gauge case.
Thus our initial focus has been on determining whether
the last condition holds.

There are two previous calculations which suggest that the center symmetry
of the single-site theory may be unbroken.
Ref.~\cite{KUY2} calculated the one-loop effective 
potential for Polyakov loops in
continuum regularization on $R^3\times S^1$, 
i.e. with one direction compactified.
A weak-coupling calculation is then 
justified if the compact direction is small 
enough. It was found that, if one uses periodic boundary conditions, 
and for massless fermions, the $Z_N$
center symmetry is unbroken for $N_f=1/2$, $1$ and $2$, but broken for $N_f=0$.
(For a recent extension to nonzero masses see Ref.~\cite{MH}.)
In words, the fermionic contributions to $V_{\rm eff}$ 
lead to a repulsion between
the eigenvalues of the Polyakov loop that overcome the attraction caused
by the gluonic contributions. The same has been found to be true in a lattice
regularized calculation~\cite{BBVeff}, although there remains some 
uncertainty over what happens if only the compact direction is 
discretized and when nontrivial forms 
of center symmetry breakings are studied~\cite{Maryland}.
These results are encouraging, but do not directly apply to the single-site
model. Perturbation theory for this theory has severe IR divergences, and the
appropriate analysis has not been done. For this reason, and because
we are interested in practice in intermediate values of the coupling 
where one-loop perturbation theory is not accurate
($b\sim 0.33$, corresponding to $\beta\sim 6$ for $N=3$),
we have chosen to do a non-perturbative calculation.

\section{The (possibly) equivalent theories}

We choose to use Wilson fermions and the Wilson gauge action. 
The (large-volume)
lattice field theory then has the action
\begin{equation}
S_{FT} =2 N \,b\,\sum_{\rm plaq} {\rm Re}{\rm Tr} U_{\rm plaq} 
+ \bar \psi \, D_{\rm W} \, \psi
\label{eq:S}
\end{equation}
where $U_{\rm plaq}$ is composed, as usual, of the fundamental irrep gauge
matrices, $b=(g^2 N)^{-1}$ is the inverse 't Hooft
coupling, and the fermion operator is ($x$ and $y$ labeling lattice sites)
\begin{equation}
\left(D_W\right)_{xy}= 
\delta_{xy} - \kappa \left[\sum_{\mu=1}^4 \left( 1 - \gamma_\mu\right) 
U^{\rm adj}_{x,\mu}\, \delta_{y,x+\mu} 
+ \left(1 + \gamma_\mu\right)U^{{\rm adj} \dag}_{x,\mu}\, \delta_{y,x-\mu}\right]\,,
\label{Dw}
\end{equation}
with the links here being in the adjoint representation.
There is a single adjoint Dirac fermion field, corresponding to $N_f=1$.
Boundary conditions are periodic in all four directions on all fields.

Notable symmetries of this theory are the usual $Z_N^4$ center symmetry and
an $SO(2)$ flavor symmetry. The latter can be seen by writing the action
in terms of two Majorana fields.

The corresponding single-site theory is obtained simply by removing the
sum over sites. One then has a theory of four link matrices $U_\mu$ and
a single $\psi$ and $\bar\psi$. The fermion operator becomes
\begin{equation}
D^{\rm red}_W=1 - \kappa \sum_{\mu=1}^4 
\left[\left( 1 - \gamma_\mu\right) U^{\rm adj}_{\mu}\, 
+ \left(1 + \gamma_\mu\right)U^{{\rm adj} \dag}_{\mu}\, \right]\,.
\end{equation}
This looks peculiar because the first derivative part connects back
(necessarily) to the same site, and so vanishes if $U_\mu=1$. This
part becomes non-trivial, however, since in the reduced model
the $U_\mu$ fluctuate around nontrivial background gauge fields, and
one can see in perturbation theory how space-time (or, more precisely,
momentum space) becomes embedded in the link matrices.

The symmetries of the single-site theory are an odd-looking gauge symmetry
[$U_\mu \to \Omega \, U_\mu \, \Omega^\dag$ for all $\mu$ with
$\Omega\in SU(N)$]
and the main player in the following---the $Z_N^4$ center symmetry:
\begin{equation}
U_\mu \to U_\mu \, z^{n_\mu} \quad {\rm with} \quad z = e^{2\pi i/N} 
\quad {\rm and}\quad n_{\mu} \in Z_N \,.
\label{ZN_sym}
\end{equation}

It is important that the putative equivalence 
applies when both theories having the same bare couplings $b$ and $\kappa$.
The continuum limit is to be taken after $N\to\infty$ so 
that (if the equivalence holds) the single-site theory has given one
information about the infinite volume field theory.

\begin{figure}[hbt]
\centerline{
\includegraphics[width=6cm]{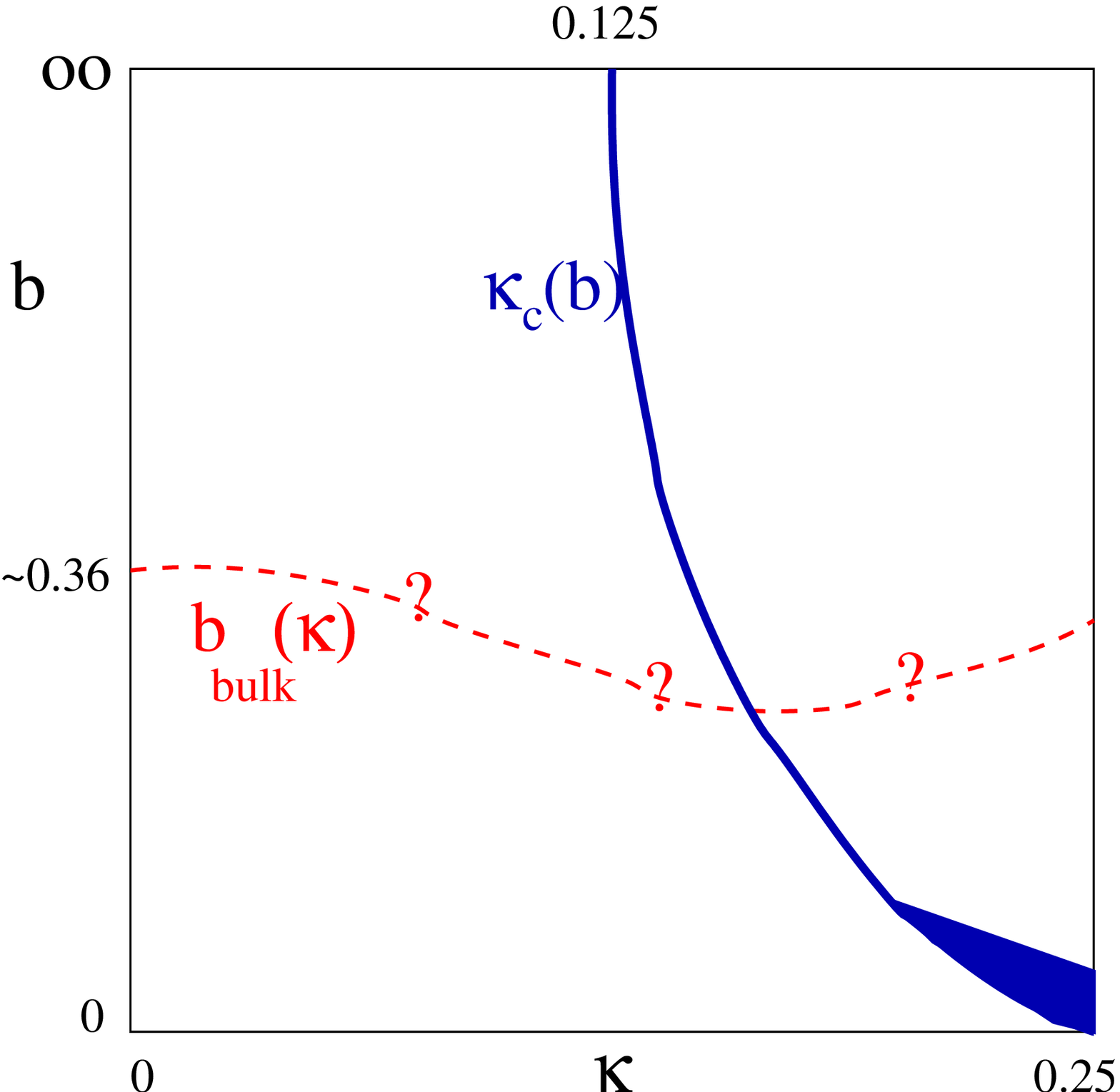}
\hspace{1cm}
\includegraphics[width=6cm]{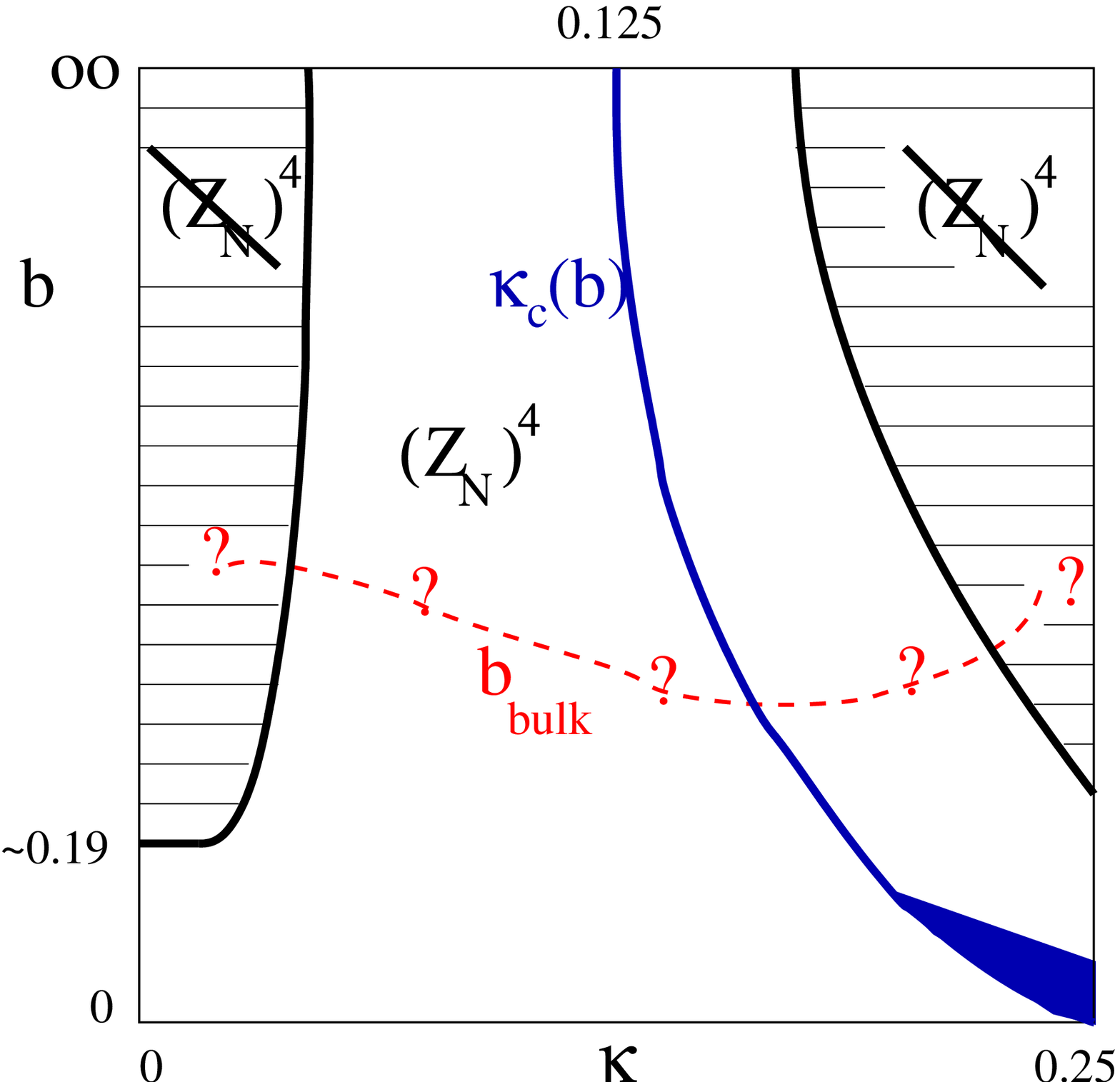}
}
\caption{Conjectured large-$N$ phase diagrams
for theories with a single adjoint Dirac fermion.
Left panel: infinite volume; right panel: single-site.
}
\label{fig:phase1}
\end{figure}

In order to test the equivalence, it is useful to have some
knowledge of the phase diagram in the $\kappa-b$ plane of the 
large-volume large-$N$ adjoint fermion theory. Since, to our knowledge, 
this theory  has not been simulated previously for any $N$, we begin by
giving an educated conjecture for the phase diagram.
This is shown in the left panel of Fig.~\ref{fig:phase1}. 
The continuum limit is at
the top while the pure gauge theory lies along the left axis.

The most important feature
is the $\kappa_c$ line, which represents the line (or region of
Aoki phase) along which (within which) the quarks attain their minimum mass.
The presence of this line/region is predicted by chiral perturbation theory
including lattice artefacts, for which the analysis is very similar
to that for QCD, except that the chiral symmetry breaking pattern is here
$SU(2)\to SO(2)$. For more details see Ref.~\cite{AEK}.
Just as in QCD, one finds that, near the continuum limit, one can either
have a first-order or Aoki-phase scenario. Our data (discussed below) 
indicate that it is the former which occurs. For very small $b$, however,
the strong-coupling expansion predicts an Aoki-phase. Thus we have
drawn the line becoming a region at small $b$.
For large $b$ the line must approach $\kappa_c=1/8$.

The other feature is the (possible) line of bulk transitions running
roughly horizontally. We know from simulations that this transition
occurs in the large-$N$ pure gauge theory, and that it is strongly
first order. It is not a deconfinement transition, but rather a lattice
artefact. This transition must extend for some distance into the plane, 
but we do not know how far. Our numerical evidence turns out to
be inconclusive as to whether the transition turns into
a crossover as one moves into the plane. In either case, we want to work on the
continuum (large $b$) side of the transition region to avoid lattice artefacts.

The right panel of Fig.~\ref{fig:phase1} shows our conjecture
for the phase diagram of the single-site model. 
In this case we must take $N\to\infty$ for there to be true transitions,
although it turns out that this is not much of an issue in practice.
The left axis (at $\kappa=0$) corresponds to the EK model, which has
the above-noted center symmetry breaking transition 
(really, a sequence of transitions) 
at $b\sim 0.19$.
Based on the perturbative calculations of Ref.~\cite{BBVeff} we
expect the symmetry to be restored when the quark becomes light,
and that it remains unbroken for a range of $\kappa$ including
$\kappa_c$. Our conjecture therefore contains a
funnel-shaped region of unbroken center symmetry in which
volume-independence should hold. We recall that, in the continuum limit, 
theories equidistant from $\kappa_c$ on the two sides are physically
equivalent, so we only need to work on one side.

\section{Some numerical results}

We have simulated the single-site model using the Metropolis
algorithm and calculating the determinant (which is real and positive)
exactly. CPU time scales as $N^8$ for a complete update of each link matrix,
and we have been able to work only up to $N=15$ using desktop PCs.
Nevertheless, this appears to be sufficient to map out the phase diagram,
since the results do not change substantially between $N=8$ and $N=15$.
For more details, see Ref.~\cite{AEK}.

We have tested our conjecture for the phase diagram by doing horizontal
hysteresis scans for couplings in the range $b=0.1-1$,
vertical scans for a variety of values of $\kappa$,
and a few high statistics runs at selected points in the putative
funnel region. Very long correlation times limit our simulations to $b\le 1$.
We note, however, that $b=1$ is a very weak
coupling, corresponding to $\beta=18$ in $SU(3)$.

We calculate the average plaquette as well as a number
of order-parameter for $Z_N^4$ symmetry-breaking: Polyakov loops 
$P_\mu={\rm Tr}(U_\mu)/N$ and ``corner loops''
$M_{\mu\nu}={\rm Tr}(U_\mu U_\nu)/N$. 
In the momentum-quenched EK model we found that one needed the $M_{\mu\nu}$ to
uncover the symmetry-breaking~\cite{QEK}, and we find the same here:
there are regions of symmetry breaking in the $\kappa>\kappa_c$ part
of the phase-plane for which
the $P_\mu$ vanish but the $M_{\mu\nu}$ do not.
To make sure that we do not miss any signals of symmetry breaking,
we have, for a few points in the plane, also calculated the 14641
different traces
\begin{equation}
K_{\vec n}\equiv {\rm Tr}\left( U^{n_1}_1\, 
U^{n_2}_2\, U^{n_3}_3\, U^{n_4}_4\right)/N\,, 
\quad {\rm with}\,\, n_\mu =0,\pm 1, \pm 2, \dots, \pm 5 \,,
\end{equation}
where $n_\mu<0$ implies hermitian conjugation. 

\begin{figure}
\centerline{
\includegraphics[width=3.2cm,height=3.2cm,angle=-90]
{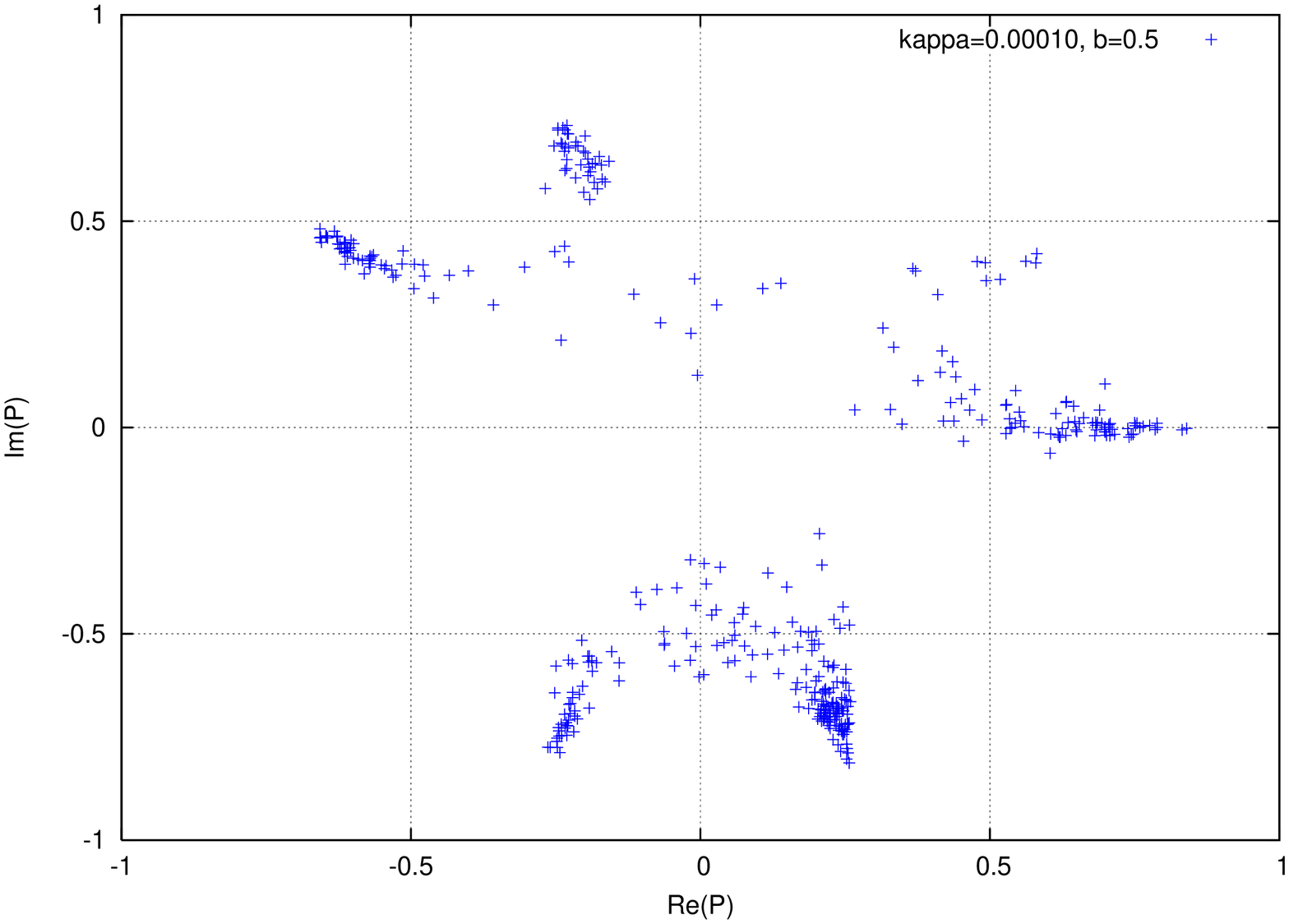}
\includegraphics[width=3.2cm,height=3.2cm,angle=-90]
{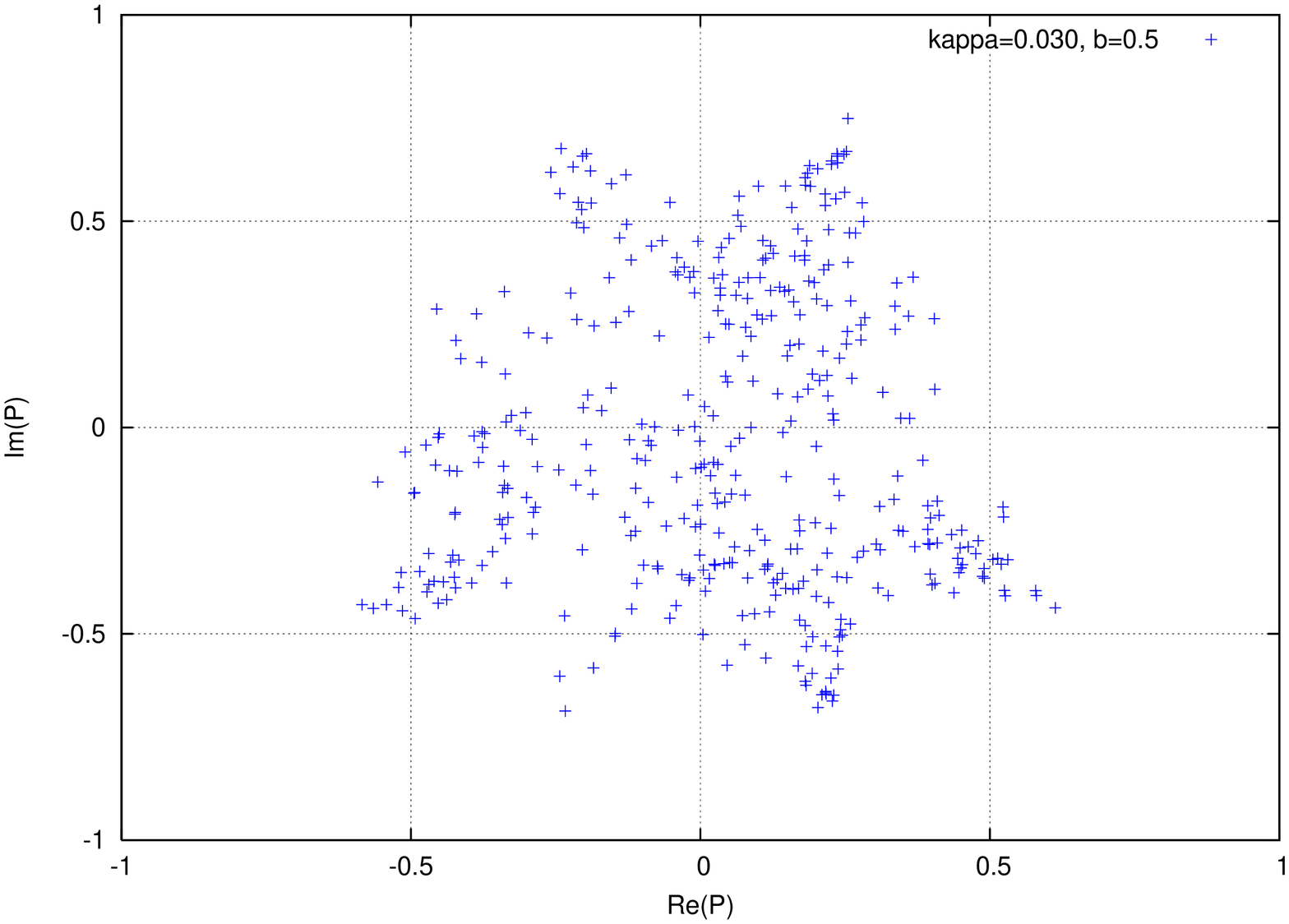}
\includegraphics[width=3.2cm,height=3.2cm,angle=-90]
{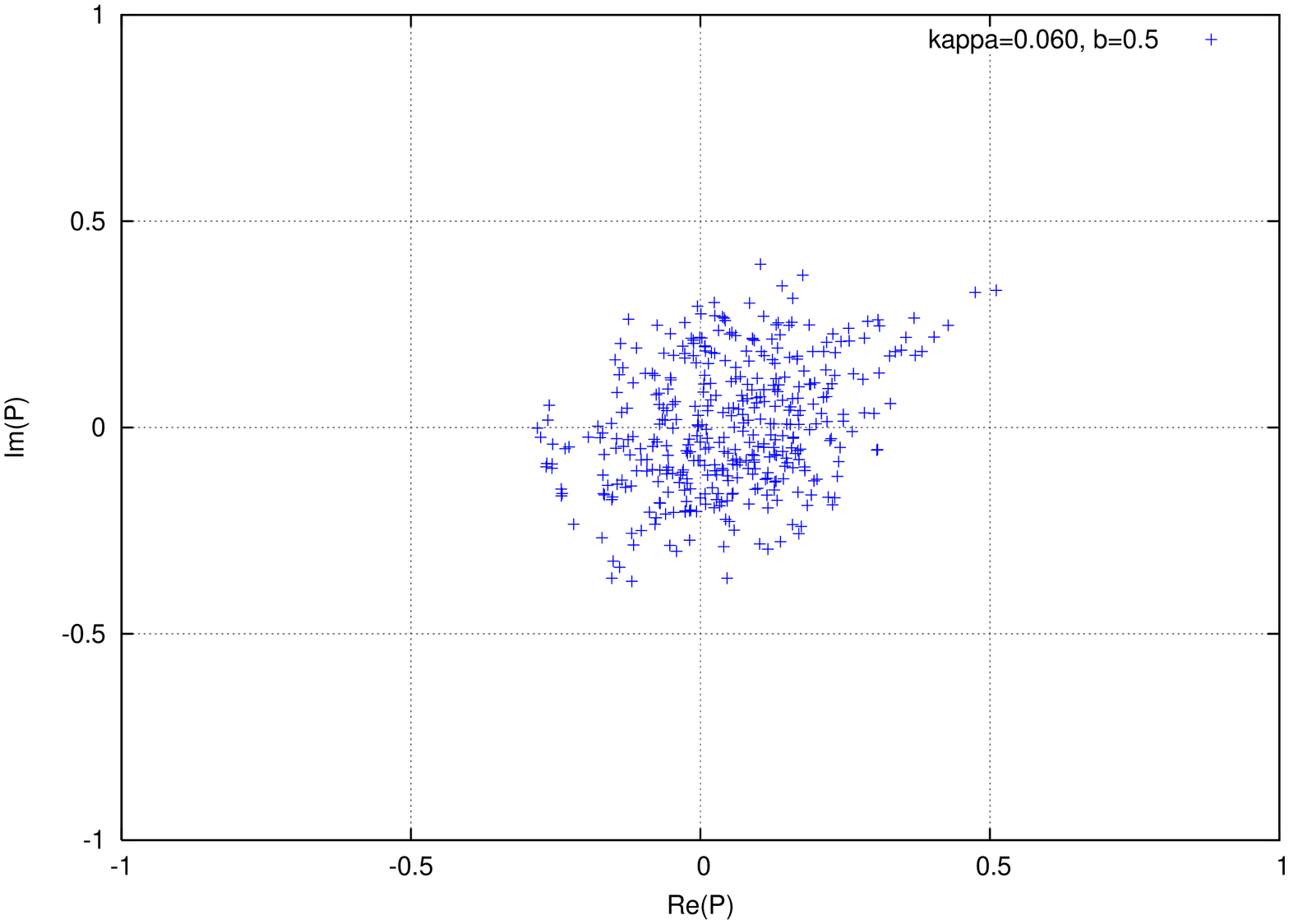}
\includegraphics[width=3.2cm,height=3.2cm,angle=-90]
{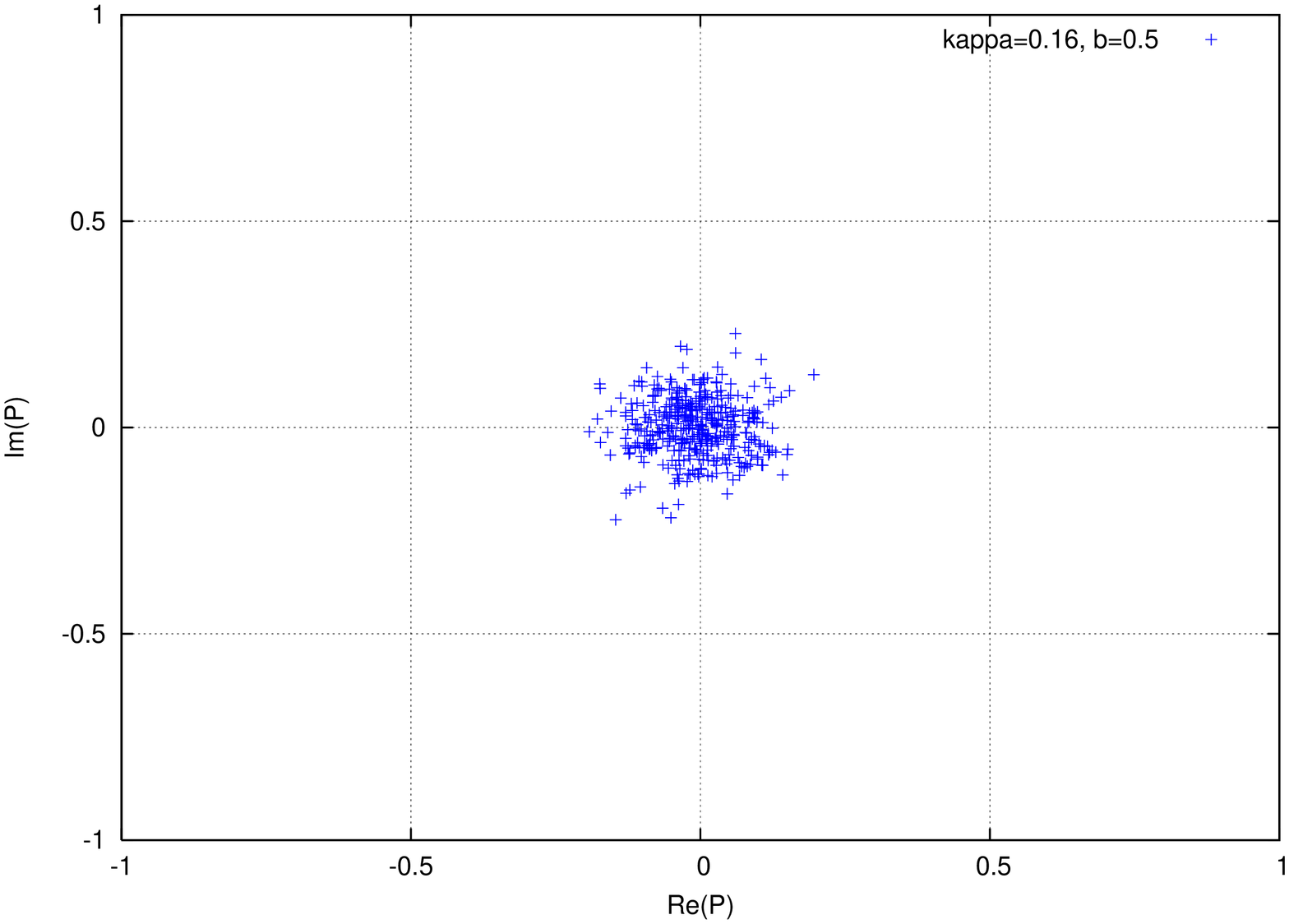}
}
\caption{Scatter plots in the complex plane for all four
Polyakov loops for $N=10$, $b=0.5$ and 
(moving from left to right)
$\kappa=0.0001$, $0.03$, $0.06$ and $0.16$.
Both horizontal and vertical scales run from $-1$ to $1$.}
\label{fig:b0.5scanN10}
\end{figure}

Our results are consistent with the conjecture described above; see
Ref.~\cite{AEK} for complete details.
As an example, we show in Fig.~\ref{fig:b0.5scanN10} scatter plots of
Polyakov loops for $N=10$ and $b=0.5$.
For $\kappa=0$, symmetry breaking is clear: all four $P_\mu$ end
up proportional to $e^{2\pi i n_\mu/10}$ (with, in this case, different
$n_\mu$), although, for this small $N$, there is still some tunneling
between different ``vacua''. Vestiges of symmetry breaking are still
apparent at $\kappa=0.03$, but are absent at $\kappa=0.06$. For this
value and for $\kappa=0.16$ (just below $\kappa_c$) the $P_\mu$ fluctuate
around the origin and average to zero within errors.
Results at $N=15$ are consistent, with the expected reduction in
tunneling for small $\kappa$ (none is seen) and the expected
reduction in the size of fluctuations.

\begin{figure}
\centerline{
\includegraphics[width=7cm,angle=-90]
{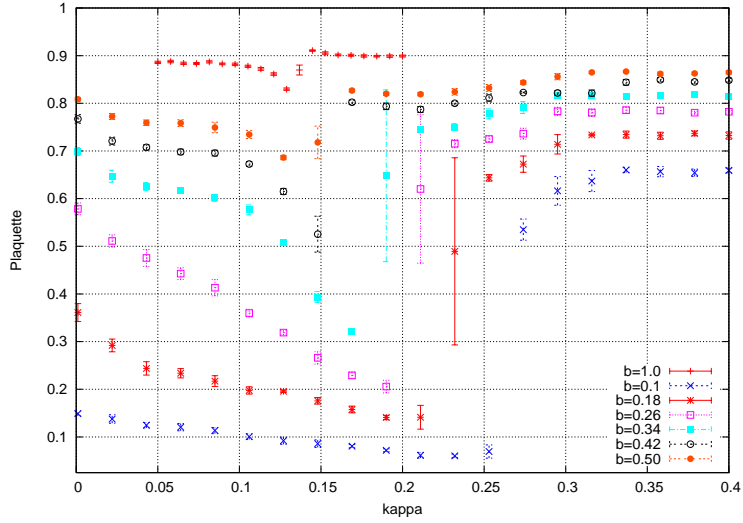}
}
\caption{Scans of the plaquette as a function of $\kappa$ for $N=8$
at $b=0.1$, $0.18$, $0.26$, $0.34$, $0.42$, $0.50$ and $1$.}
\label{fig:firstorder}
\end{figure}

Evidence for the presence
of a first-order  transition is shown in Fig.~\ref{fig:firstorder}:
there is a jump in the average plaquette for each value of $b$.
We also observe hysteresis at this ``transition'', and a consistent
behavior for $N=8-15$. The transition is not, however, associated with center
symmetry breaking and we interpret it as an example of
the first-order scenario predicted by chiral perturbation theory. 
This is supported by two other observations. First,
its position is consistent with the theoretical expectations---moving 
from $1/4$ at strong coupling to $1/8$ at weak coupling.
Second, the jump at the transition decreases as $b$ increases,
consistent with the theoretical expectation that the transition
is driven by lattice artefacts.

\section{Conclusions and Outlook}

In summary, our evidence to date is consistent with the conjecture
in the right panel of Fig.~\ref{fig:phase1}, including the
presence of a window around $\kappa_c$ of finite width in which
the $Z_N^4$ symmetry is unbroken.
Accepting this result, the adjoint $N_f=1$ theory provides the
first successful example of reduction to a single-site.
Furthermore, it appears that the funnel region is broad
enough to allow one to work with quarks with masses
$m\gg \Lambda_{\rm QCD}$ (and possibly with $m\sim 1/a$)
so that the long-distance physics is that of the pure gauge theory.
In this case we have, in effect, a version of the original EK model
in which the effect of the adjoint quarks is to maintain the
repulsion among eigenvalues of the Polyakov loop,
presumably by inducing a tower of double-trace operators
similar to that suggested in Ref.~\cite{DEK} as a cure for the EK model.
If, on the other hand, we work near $\kappa_c$ we have a single site
model that lies ``within $1/N$'' of physical two-flavor QCD.

Our analysis can be extended in several ways. Our interpretation of
the critical line can be checked by calculating the pion mass. We can
also calculate other  meson masses and the string tension, 
and the distribution of the
eigenvalues of Dirac operators for various valence fermions.
Work in these directions is underway using the configurations in hand.
We also plan on speeding up the algorithm so as to move to larger values
of $N$.

Finally, we note that another recent lattice study 
(using overlap as opposed to our Wilson fermions) finds 
preliminary evidence for a failure of reduction~\cite{Hietanen}.
This discrepancy with our findings will clearly need to be resolved.

\end{document}